\documentclass[aps,prl,preprintnumbers,showpacs,nofootinbib,floatfix]{revtex4}
\usepackage{graphicx}
\usepackage{comment}
\usepackage{dcolumn}
\usepackage{bm}
\begin{document}
\newcommand{\newc}{\newcommand}
\newc{\ra}{\rightarrow}
\newc{\lra}{\leftrightarrow}
\newc{\beq}{\begin{equation}}
\newc{\eeq}{\end{equation}}
\newc{\barr}{\begin{eqnarray}}
\newc{\earr}{\end{eqnarray}}
\newcommand{\Od}{{\cal O}}
\newcommand{\lsim}   {\mathrel{\mathop{\kern 0pt \rlap
  {\raise.2ex\hbox{$<$}}}
  \lower.9ex\hbox{\kern-.190em $\sim$}}}
\newcommand{\gsim}   {\mathrel{\mathop{\kern 0pt \rlap
  {\raise.2ex\hbox{$>$}}}
  \lower.9ex\hbox{\kern-.190em $\sim$}}}

\title{CAN SOLAR NEUTRINOS BE A SERIOUS BACKGROUND IN DIRECT DARK MATTER SEARCHES?
 }
\author{J. D. Vergados$^{(1),(2)}$\thanks{Vergados@uoi.gr} and  H. Ejiri$^{(3),(4),(5)}$\thanks{ejiri@rcnp.osaka-u.ac.jp}}
\affiliation{$^{(1)}${\it Cyprus  University  of Technology (CUT), P.O. Box 50329, 3603 Limassol, Cyprus,}}
\affiliation{$^{(2)}${\it Physics Department, University of Ioannina, Ioannina, GR 451 10, Greece,}}
\affiliation{$^{(3)}${\it RCNP, Osaka University, Osaka, 567-0047, Japan,}}
\affiliation{$^{(4)}${\it National Institute of Radiological Sciences, Chiba,263-8555,Japan}}
\affiliation{$^{(5)}${\it Nuclear Science, Czech Technical University, Brehova, Prague, Czech Republic.}}
\begin{abstract}
The coherent contribution of all neutrons in neutrino nucleus scattering due to the neutral current is examined considering the boron solar neutrinos. These neutrinos could potentially become a source of background in the future dark matter searches aiming at nucleon cross sections in the region well below the $10^{-10}$pb, i.e a few events per ton per year. 
\end{abstract}
\date{\today}
\maketitle
\section{Introduction.}
The universe is observed to contain large amounts of dark matter
\cite{SPERGEL,WMAP06}, and its contribution to the total energy density is
estimated to be $\sim 25\%$. This non-baryonic dark matter component,
responsible for the growth of cosmological perturbations through
gravitational instability, has still not been detected directly. Even
though there exists firm indirect evidence from the halos of dark matter in galaxies
and clusters of galaxies it is
essential to detect matter directly.

The possibility of direct detection, however, depends on the nature of
the dark matter constituents, i.e the WIMPs (weakly interacting massive particles).
 Supersymmetry naturally provides
candidates for these constituents
\cite{goodwit,KVprd,ellrosz,ref2,EOSS05}. In the most favored scenario of
supersymmetry, the lightest supersymmetric particle (LSP) can be
described as a Majorana fermion, a linear combination of the neutral
components of the gauginos and higgsinos. Other possibilities also exist,
see, e.g. some models in universal theories with extra dimensions \cite{SERVANT}.
Since the  WIMPs are expected to be very massive
($m_{{WIMP}} \geq  30$ GeV) and extremely non-relativistic with
average kinetic energy $ \langle T \rangle \simeq$ 50 keV
$\left(m_{{WIMP}}/ 100\, {\rm GeV} \right)$, a WIMP interaction
with a nucleus in an underground detector is not likely to produce
excitation.  As a result, WIMPs can be directly detected mainly via
the recoil of a nucleus ($A,Z$) in elastic scattering. The event rate
for such a process can be computed from the following ingredients:
\begin{enumerate}
\item An effective Lagrangian at the elementary particle (quark)
level obtained in the framework of the prevailing particle theory. For supersymmetry 
this is achieved as described, e. g., in refs. \cite{ref2,JDV96}.  For 
Kaluza Klein theories in universal extra dimensions see, e. g., some recent calculations
 \cite{OikVerMou}. 
Invariably this ingredient is the most important element, but at present,  unfortunately, with
the greatest uncertainty in getting the event rate, especially since the WIMP mass is
quite uncertain.
\item A well defined procedure for transforming the amplitude
obtained using the previous effective Lagrangian from the quark to
the nucleon level, i.e. a quark model for the nucleon. This step
in SUSY models is non-trivial, since the obtained results depend crucially on the
content of the nucleon in quarks other than u and d.
\item Knowledge of the relevant nuclear matrix elements
\cite{Ress,DIVA00}, obtained with reliable many-body nuclear wave functions. 
Fortunately, in the case of the scalar
coupling, which is viewed as the most important, the situation is
a bit simpler, as only the nuclear form
factor is needed.
\item Knowledge of the WIMP density in our vicinity and its velocity
distribution. Since the essential input here comes from the rotation
curves, dark matter candidates other than the LSP are
also characterized by similar parameters. One does not know for sure
what these parameters are. The most common models are isothermal Maxwell-Boltzmann (M-B) distributions
\cite{Druk,Verg00,evans00,fornen}, anisotropic velocity distributions
described by Tsallis type functions \cite{HANSEN06,Host,rocco,VerHanH}, i.e. functions which in some limit lead
to the M-B  distributions, and variants of the M-B distributions arising when dark matter is coupled to dark energy 
\cite{TETRVER06}. Non isothermal models, like those arising in the Eddington approach have also been
considered \cite{EDDIN,UK01,VEROW06,BCFS02}. These various models give similar (time averaged) rates and differ
only in the predictions regarding the modulation. So one may assume that the uncertainties in this case are quite
small. 
\end{enumerate}

The particle physics in conjunction with the structure of the nucleon provide the nucleon 
cross sections. Since, as we have already mentioned,  the particle physics parameters most likely will result in very small cross sections,
the most ambitious future dark matter experiments like the XENON-ZEPLIN aim at detecting 10 events per ton per year. At this  level one may encounter very bothersome backgrounds. One such background may come from  the high energy boron solar neutrinos (the other neutrinos are characterized either by too small energy or much lower fluxes).

 During the last years various detectors aiming at detecting recoiling nuclei have been
 developed 
in connection with  dark matter 
searches \cite{CDMALL} with thresholds in the few keV region. Recently, however, it has become feasible to detect neutrinos by measuring
 the recoiling nucleus  and employing gaseous detectors with much lower threshold energies \cite{VERGIOM06}. Thus
 one is able to explore the advantages offered by the neutral current interaction, exploring ideas put forward more than a decade  ago \cite{DKLEIN}. 
Furthermore this interaction, through its vector component, can lead
 to coherence, i.e. an additive contribution of all neutrons in the nucleus (the vector  contribution of the
 protons is tiny, so the coherence is mainly due to the neutrons of the nucleus).
 
 In this paper we will derive the differential neutrino nucleus cross section and the associated event rate
for the elastic (coherent) neutrino-nucleus scattering. Then
 we will utilize the available information regarding the energy spectrum of solar boron neutrinos and estimate
 the expected number of events for light as well as heavy nuclear target.  Finally we will compare 
 the recoil spectrum  and total event rate
associated with WIMPs with  that due to neutrinos.

  \section{A brief discussion of the rates for direct WIMP detection}
   Before proceeding with the evaluation of the event rate for nuclear recoils originating from
   the neutrino nuclear scattering we will briefly discuss the WIMP-nucleus recoiling events.
   We begin by saying that
the shape of the differential event rate for WIMP detection cannot be precisely estimated, since,
as we mentioned in the introduction, the WIMP-nucleon 
 cross section is not known. Especially its dependence  on the WIMP mass is not  known.
 The non-directional differential rate folded with the WIMP velocity distribution is given 
by \cite{Verg01,JDV03,JDVSPIN04,JDV06}:
\begin{equation}
 \Big<\frac{dR}{du}\Big> =
\frac{\rho (0)}{m_{\chi}} \frac{m}{Am_N} \sqrt{\langle
\upsilon^2\rangle }\int
           \frac{   |{\boldmath \upsilon}|}
{\sqrt{ \langle \upsilon^2 \rangle}}
 f(\mbox{\boldmath $\upsilon$},\mbox{\boldmath $\upsilon$}_E)
                       \frac{d\sigma (u,\upsilon )}{du} d^3
 \mbox{\boldmath $\upsilon$}
\label{3.11}
\end{equation}
where
$f(\mbox{\boldmath $\upsilon$},\mbox{\boldmath $\upsilon$}_E)$ essentially is the WIMP velocity distribution
in the laboratory frame with $\mbox{\boldmath $\upsilon$}_E$ the velocity of the Earth.

The differential cross section is given by:
\beq
d\sigma (u,\upsilon)== \frac{du}{2 (\mu _r b\upsilon )^2}
 [(\bar{\Sigma} _{S}F(u)^2
                       +\bar{\Sigma} _{spin} F_{11}(u)] 
\label{2.9}
\end{equation}
where $ u$ the energy transfer $Q$ in dimensionless units given by
\begin{equation}
 u=\frac{Q}{Q_0}~~,~~Q_{0}=[m_pAb]^{-2}=40A^{-4/3}~MeV
\label{defineu}
\end{equation}
 with  $b$ is the nuclear (harmonic oscillator) size parameter. $F(u)$ is the
nuclear form factor and $F_{11}(u)$ is the spin response function associated with
the isovector channel.

The scalar
cross section is given by:
\begin{equation}
\bar{\Sigma} _S  =  (\frac{\mu_r(A)}{\mu_r(p)})^2
                           \sigma^{S}_{p,\chi^0} A^2
 \left [\frac{1+\frac{f^1_S}{f^0_S}\frac{2Z-A}{A}}{1+\frac{f^1_S}{f^0_S}}\right]^2
\approx  \sigma^{S}_{N,\chi^0} (\frac{\mu_r(A)}{\mu_r (p)})^2 A^2
\label{2.10}
\end{equation}
(since the heavy quarks dominate the isovector contribution is
negligible). $\sigma^S_{N,\chi^0}$ is the LSP-nucleon scalar cross section.
The spin cross section is given by:
\begin{equation}
\bar{\Sigma} _{spin}  =  (\frac{\mu_r(A)}{\mu_r(p)})^2
                           \sigma^{spin}_{p,\chi^0}~\zeta_{spin}
\label{2.10a}
\end{equation}
$~\zeta_{spin}$ the nuclear spin ME.

Integrating over the energy transfer $u$ we obtain the  event rate for the coherent WIMP-nucleus elastic scattering, which is given by \cite{Verg01,JDV03,JDVSPIN04,JDV06}:
\barr
R&=& \frac{\rho (0)}{m_{\chi^0}} \frac{m}{m_p}~
              \sqrt{\langle v^2 \rangle } 
\nonumber\\
& &\left [f_{coh}(A,\mu_r(A)) \sigma_{p,\chi^0}^{S}+f_{spin}(A,\mu_r(A))\sigma _{p,\chi^0}^{spin}~\zeta_{spin} \right]
\label{fullrate}
\earr
with
\beq
f_{coh}(A, \mu_r(A))=\frac{100\mbox{GeV}}{m_{\chi^0}}\left[ \frac{\mu_r(A)}{\mu_r(p)} \right]^2 A~t_{coh}\left(1+h_{coh}cos\alpha \right)
\eeq
\beq
f_{spin}(A, \mu_r(A))=\left[ \frac{\mu_r(A)}{\mu_r(p)} \right]^2 \frac{t_{spin}(A)}{A}
\eeq
 In this work we will not be concerned with the spin cross section. The
parameter $t_{coh}$ takes into account the folding of the nuclear form factor with the WIMP
 velocity distribution, $h_{coh}$ deals with the modulation due to the motion of the Earth and $\alpha$ is
 the phase of the Earth (zero around June 2nd).

\begin{figure}[!ht]
 \begin{center}
\rotatebox{90}{\hspace{-0.0cm} { $\frac{dt_{coh}}{du} \longrightarrow$}}
\includegraphics[scale=1.1]{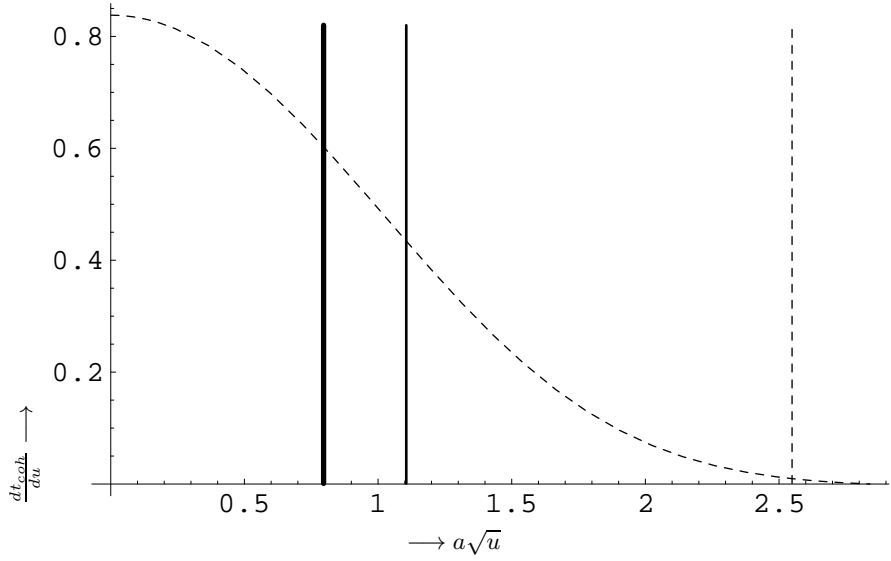}\\
{\hspace{0.0cm} $\longrightarrow  a \sqrt{u}$}
 \caption{The differential event rate for the coherent process,
in arbitrary units, as a function of the parameter $a\sqrt{u}$, where $a=1.4 10^{-3}(\mu_r(A)c^2b/(\hbar c) $ with $\mu_r(A)$ the WIMP-nucleus reduced mass and $b$ the size parameter for the nucleus. $u $ is essentially the energy 
transfer $Q$, $u=Q/Q_0$, $Q_0=4.1 10^4 A^{-4/3}$.
Due to the nuclear form factor not all the range of $u$ is exploitable in direct WIMP detection. For $^{131}$Xe, e.g., effectively there is
 an upper cut off value indicated by dotted line, fine line and thick line for a WIMP mass of $30$, $100$ and $200$ GeV respectively. Any lower cut off is due to the threshold.}
 \label{fig:dtdu}
  \end{center}
  \end{figure}

\begin{figure}[!ht]
 \begin{center}
\rotatebox{90}{\hspace{-0.0cm} {$\frac{dR_{coh}}{dQ}$  $\rightarrow keV^{-1} \frac{\sigma_{p,\chi^0}^{S}}{10^{-6}pb}$}}
\includegraphics[scale=1.1]{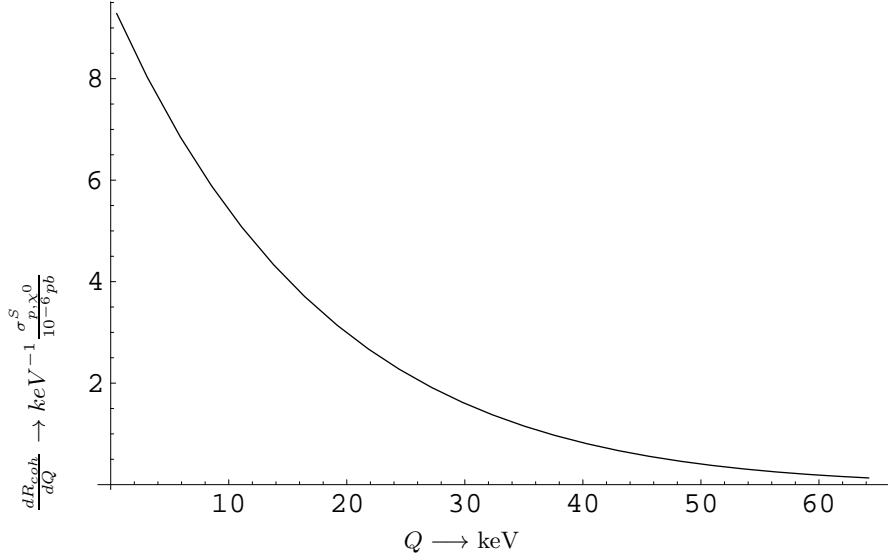}\\
{\hspace{0.0cm} $Q \longrightarrow$  keV}
 \caption{We  show the differential event rate  $ {dR_{coh}}/{dQ}$ for the coherent process , as a function of the energy transfer,
   for  a WIMP mass, $m_{\chi}$, of 100 GeV
in the case of $^{131}$Xe.}
\label{fig:drdQ127}
   \end{center}
  \end{figure}

\begin{figure}[!ht]
 \begin{center}
\rotatebox{90}{\hspace{-0.0cm} {Event rate per kg-yr
$\rightarrow \frac{\sigma_{p,\chi^0}^{S}}{10^{-6}pb}$}}
\includegraphics[scale=1.0]{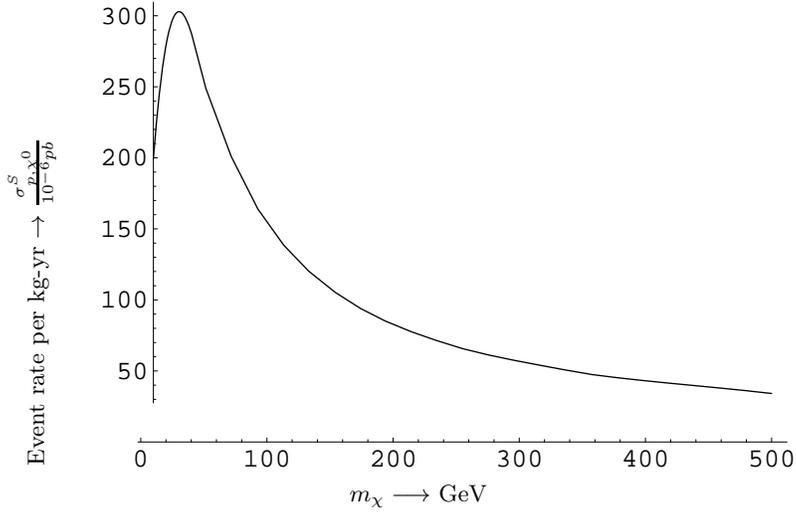}\\
{\hspace{0.0cm} $m_{\chi} \longrightarrow$ GeV }
 \caption{We  show the total event rate   for the coherent process  as a function of the  WIMP mass in the case of $^{131}$Xe for zero threshold.}
 \label{fig:rate131}
   \end{center}
  \end{figure}

 The number of events in time $t$ due to the scalar interaction, which leads to coherence  \cite{TETRVER06}, is:
\barr
 R&&\simeq  1.60~10^{-3}\times 
 \nonumber\\
 &&
\frac{t}{1 \mbox{y}} \frac{\rho(0)}{ \mbox {0.3GeVcm}^{-3}}
\frac{m}{\mbox{1Kg}}\frac{ \sqrt{\langle
v^2 \rangle }}{280 {\mbox {km s}}^{-1}}\frac{\sigma_{p,\chi^0}^{S}}{10^{-6} \mbox{ pb}} f_{coh}(A, \mu_r(A))
\label{eventrate}
\earr
  Assuming a constant nucleon cross section we get the differential rate 
 indicated by Figs \ref{fig:dtdu} and \ref{fig:drdQ127}.
   This shape was obtained in the coherent mode, but we expect to have a similar shape due to the spin
   \cite{JELLIS,JDVSPIN04}. The essential difference is that one needs  as input
a much larger nucleon cross section due to the spin.  Indeed
   for a heavy target the spin mode has no chance, if the nucleon cross section
   due to the spin is of the same magnitude with that associated with the coherent mode.\\
   The total (time averaged) coherent event rate is shown in 
Figs \ref{fig:rate131} and \ref{fig:rate32_abs}.

      \begin{figure}[!ht]
 \begin{center}
\rotatebox{90}{\hspace{-0.0cm} {$\frac{dR_{coh}}{dQ}$  $\rightarrow keV^{-1} \frac{\sigma_{p,\chi^0}^{S}}{10^{-6}pb}$}}
\includegraphics[scale=1.1]{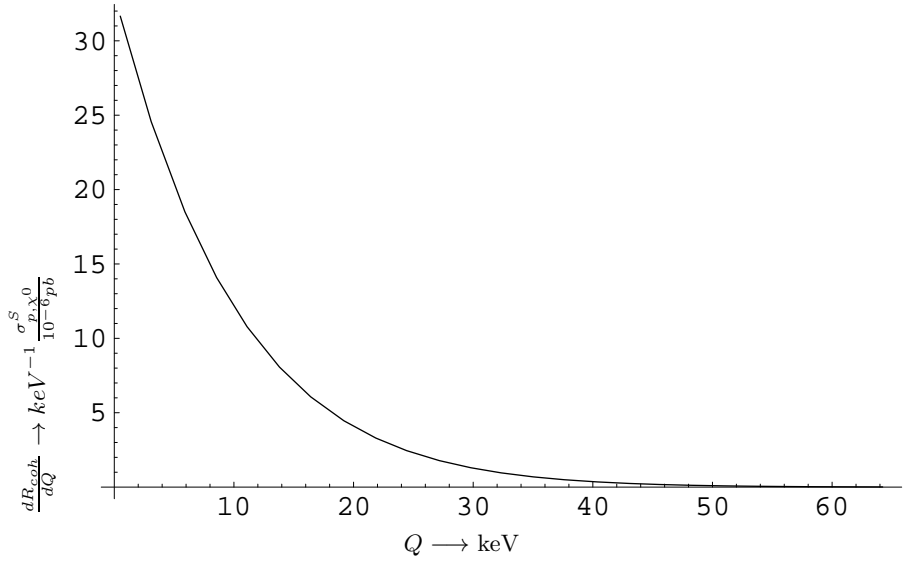}\\
{\hspace{0.0cm} $Q \longrightarrow$  keV}
 \caption{The same as in Fig. \ref{fig:drdQ127} a WIMP mass of 30 GeV.}
\label{fig:drdQ_30_127}
   \end{center}
  \end{figure}
        \begin{figure}[!ht]
 \begin{center}
\rotatebox{90}{\hspace{-0.0cm} {$\frac{dR_{coh}}{dQ}$  $\rightarrow keV^{-1} \frac{\sigma_{p,\chi^0}^{S}}{10^{-6}pb}$}}
\includegraphics[scale=1.1]{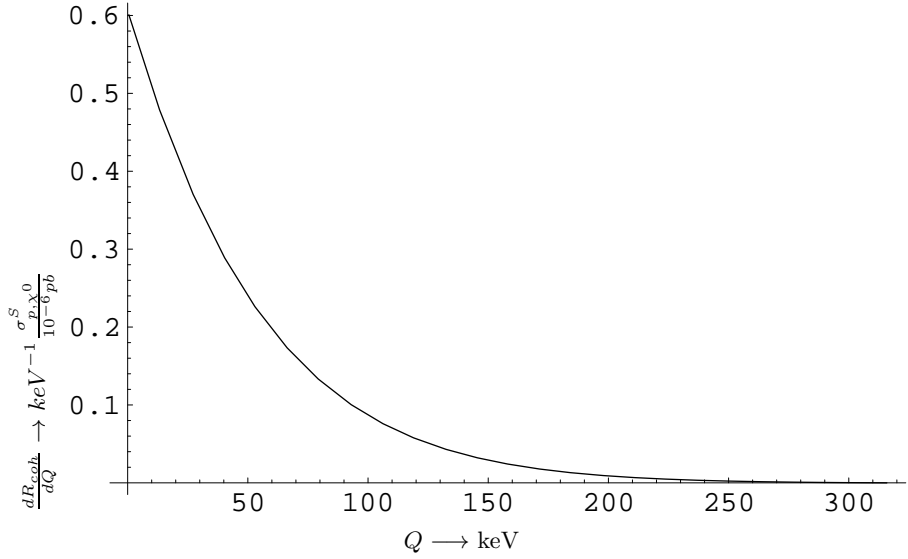}\\
{\hspace{0.0cm} $Q \longrightarrow$  keV}
 \caption{The same as in Fig. \ref{fig:drdQ127} in the case of the A=32 target.}
\label{fig:drdQ_100_32}
   \end{center}
  \end{figure}
        \begin{figure}[!ht]
 \begin{center}
\rotatebox{90}{\hspace{-0.0cm} {Event rate per kg-yr
$\rightarrow \frac{\sigma_{p,\chi^0}^{S}}{10^{-6}pb}$}}
\includegraphics[scale=1.0]{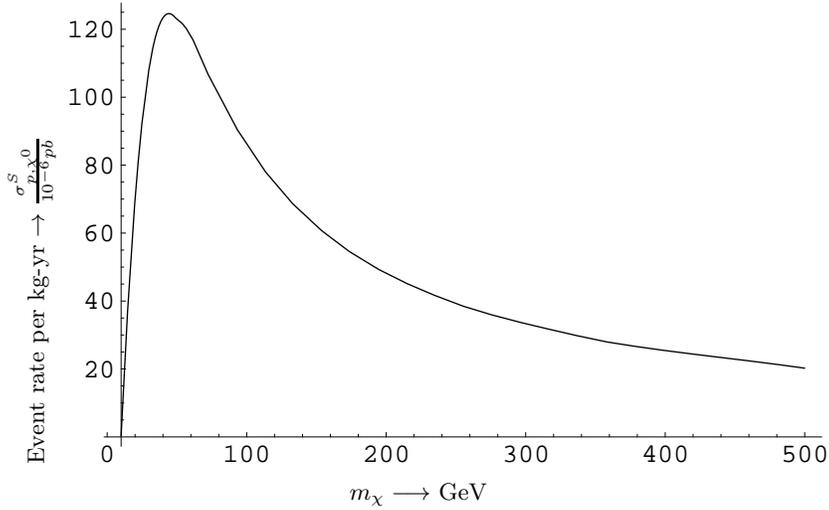}\\
{\hspace{0.0cm} $m_{\chi} \longrightarrow$ GeV }
 \caption{The same as in Fig. \ref{fig:rate131} for a detector energy threshold of 10 keV.}
 \label{fig:rate131_10}
   \end{center}
  \end{figure}
        \begin{figure}[!ht]
 \begin{center}
\rotatebox{90}{\hspace{-0.0cm} {Event rate $\rightarrow$ kg-yr
}}
\includegraphics[scale=1.1]{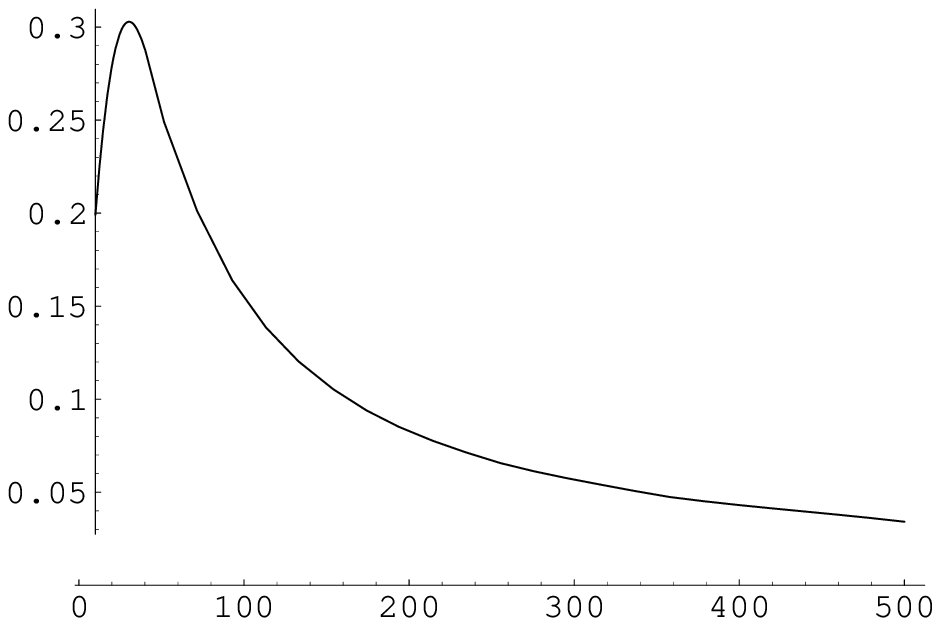}\\
{\hspace{0.0cm} $m_{\chi} \longrightarrow$ GeV }
 \caption{We  show the total event rate   for the coherent process  as a function of the  WIMP mass in the case of $^{131}$Xe, employing a nucleon cross section of $\sigma_{p,\chi^0}^{S}=10^{-9}pb $. }
\label{fig:rate131_abs}
   \end{center}
  \end{figure}
   \begin{figure}[!ht]
 \begin{center}
\rotatebox{90}{\hspace{-0.0cm} {Event rate $\rightarrow$ kg-yr
}}
\includegraphics[scale=1.1]{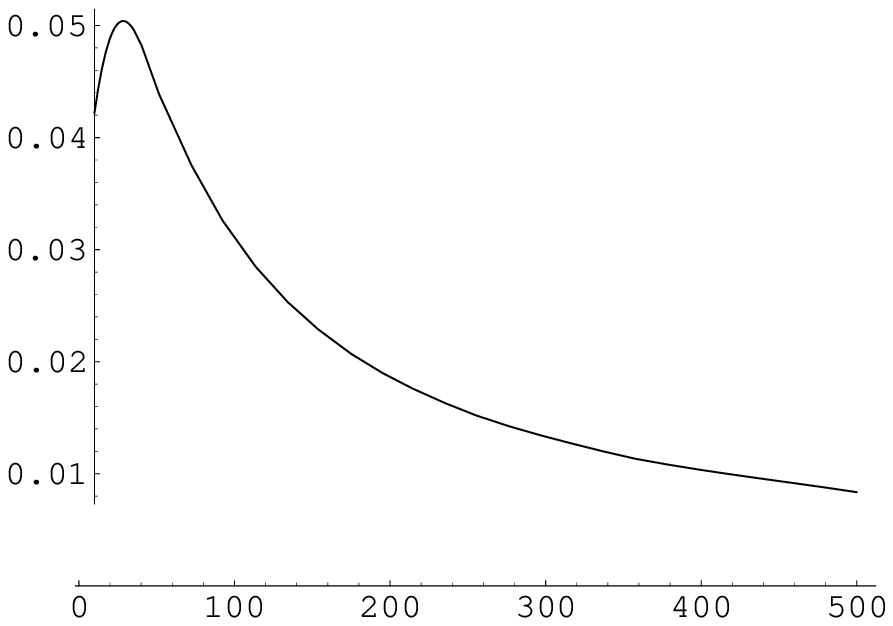}\\
{\hspace{0.0cm} $m_{\chi} \longrightarrow$ GeV }
 \caption{The same as in Fig. \ref{fig:rate131_abs} in the case of the A=32 system.}
\label{fig:rate32_abs}
   \end{center}
  \end{figure}
\section{ Elastic Neutrino nucleon Scattering}
The cross section for elastic neutrino nucleon scattering has extensively been studied.
It has been shown that at low energies it can be simplified and  be cast in the form:
\cite{BEACFARVOG},\cite{VogEng}:
 \begin{eqnarray}
 \left(\frac{d\sigma}{dT_N}\right)_{weak}&=&\frac{G^2_F m_N}{2 \pi}
 [(g_V+g_A)^2\\
\nonumber
&+& (g_V-g_A)^2 [1-\frac{T_N}{E_{\nu}}]^2
+ (g_A^2-g_V^2)\frac{m_NT_N}{E^2_{\nu}}]
 \label{elasw}
  \end{eqnarray}
  where $m_N$ is the nucleon mass and $g_V$, $g_A$ are the weak coupling constants. Neglecting their
  dependence on the momentum transfer to the nucleon they take the form:
  \beq
 g_V=-2\sin^2\theta_W+1/2\approx 0.04~,~g_A=\frac{1.27}{2} ~~,~~(\nu,p)
\label{gcoup1}
\eeq
\beq
g_V=-1/2~,~g_A=-\frac{1.27}{2}~~,~~(\nu,n)
 \label{gcoup2}
 \eeq
 In the above expressions for the axial current the renormalization
in going from the quark to the nucleon level was taken into account. For antineutrinos $g_A\rightarrow-g_A$.
To set the scale we write:
\beq \frac{G^2_F m_N}{2 \pi}=5.14\times 10^{-41}~\frac{\mbox{cm}^2}{\mbox{MeV}}
\label{weekval} 
\eeq
The nucleon energy depends on the  neutrino energy and the
scattering angle, the angle between the direction of the recoiling particle and that of the incident neutrino. In the laboratory frame  it is given by:
\beq
T_N= \frac{2~m_N (E_{\nu}\cos{\theta})^2}{(m_N+E_\nu)^2-(E_{\nu}
\cos{\theta})^2}~~,~~0\leq \theta\leq \pi/2
\label{eq:TN}
\eeq
(forward scattering). For sufficiently small neutrino energies, the last equation can be simplified as follows:
$$ T_N \approx \frac{ 2(E_\nu \cos{\theta})^2}{m_N}$$
The above formula can be generalized to any target and can be written in dimensionless form
as follows:
\beq
y=\frac{2\cos^2{\theta}}{(1+1/x_{\nu})^2-\cos^2{\theta}}~~,~~
y=\frac{T_{recoil}}{m_{recoil}},x_{\nu}=\frac{E_{\nu}}{m_{recoil}}
\label{recoilen}
\eeq 
In the present calculation we will treat $x_{\nu}$ and $y$ as dynamical variables, in line with CDM recoils. One, of course,
equally well could have chosen $x_{\nu}$ and $\theta$ as relevant variables.

 The maximum  energy occurs when $\theta=0$, i.e.:
\beq
 y_{max}=\frac{2}{(1+1/x_{\nu})^2-1},
 \label{Eq:ymax}
 \eeq
in agreement with Eq. (2.5) of ref. \cite{BEACFARVOG}.
  One can invert Eq. \ref{recoilen} and get the  neutrino energy associated with a given recoil energy and
scattering angle. One finds
\beq
  x_{\nu}=\left[-1+\cos{\theta} \sqrt{1+\frac{2}{y}} \right]^{-1}~~,~~0\leq \theta\leq \pi/2
  \label{xofyxi}
  \eeq
  The minimum neutrino energy for a given recoil energy is given by:
    \beq
  x^{min}_{\nu}=\left[-1+\sqrt{1+\frac{2}{y}} \right]^{-1}=\frac{y}{2}(1+\sqrt{1+\frac{2}{y}})
  \label{xofy}
  \eeq
 in agreement with Eq. (4.2) of ref. \cite{BEACFARVOG}. The last equation is useful in obtaining the differential cross section (with respect to the recoil energy) after folding with the neutrino spectrum
\section {Coherent neutrino nucleus scattering}
\label{quenching}
From the above expressions we see that the vector current contribution, which may lead to coherence, is negligible
in the case of the protons. Thus the coherent contribution \cite{PASCHOS} may come from the neutrons and is expected to be
proportional to the square of the neutron number.
The neutrino-nucleus scattering can be derived in analogous fashion. It can also be obtained from the amplitude of the neutrino nucleon scattering 
by  employing the appropriate kinematics, i.e. those involving the elastically scattered nucleus and
 the substitution 
$${\bf q}\Rightarrow \frac{{\bf p}}{A}~~,~~E_N \Rightarrow \sqrt{m_N^2+\frac{{\bf p}^2}{A^2}}=\frac{E_A}{A}$$
with ${\bf q}$ the nucleon momentum and ${\bf p}$ the nuclear momentum.  
Under the above assumptions the neutrino-nucleus cross section takes the form:
 \begin{eqnarray}
 \left(\frac{d\sigma}{dT_A}\right)&=&\frac{G^2_F Am_N}{2 \pi}
 [(M_V+M_A)^2 \left (1+\frac{T_A}{E_{\nu}} \right )
 \nonumber\\
&+ &(M_V-M_A)^2 
(1-\frac{T_A}{E_{\nu}})^2
+ (M_A^2-M_V^2)\frac{Am_NT_A}{E^2_{\nu}} ]
 \label{elaswA}
  \end{eqnarray}
  where $M_V$ and $M_A$ are the nuclear matrix elements associated with the vector and the axial currents
  respectively and $T_A$ is the energy of the recoiling nucleus.
 The axial current contribution vanishes for $0^+ \Rightarrow 0^+$ transitions. Anyway it is negligible
  in front of the coherent scattering due to neutrons. Thus the previous formula is reduced to:
\beq
 \left(\frac{d\sigma}{dT_A}\right)_{weak}=\frac{G^2_F Am_N}{2 \pi}~(N^2/4) F_{coh}(T_A,E_{\nu}),
 \label{elaswAV1}
\eeq
with
\beq
F_{coh}(T_A,E_{\nu})= F^2(q^2)
  \left ( 1+(1-\frac{T_A}{E_{\nu}})^2
-\frac{Am_NT_A}{E^2_{\nu}} \right) 
 \label{elaswAV2}
  \eeq
  where $F(q^2)= F(T_A^2+2 A m_N T_A)$ is the nuclear form factor.
  
   The nuclear form factor makes a sizable contribution in the case of the more energetic supernova
neutrinos. In the case of solar neutrinos, due to the small nuclear recoil energy,
 the form factor is expected to play a minor role numerically.  
   
\begin{figure}[!ht]
 \begin{center}
 \rotatebox{90}{\hspace{1.0cm} {$f_{coh}(T_A) \rightarrow $}}
\includegraphics[scale=0.6]{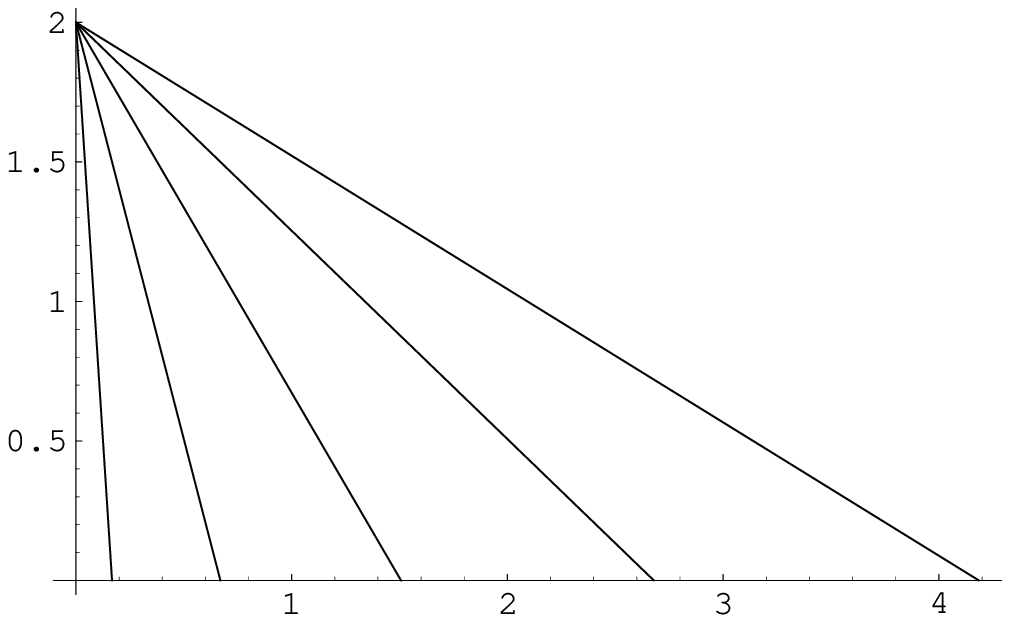}
 \rotatebox{90}{\hspace{1.0cm} {$f_{coh}(y) \rightarrow $}}
\includegraphics[scale=0.6]{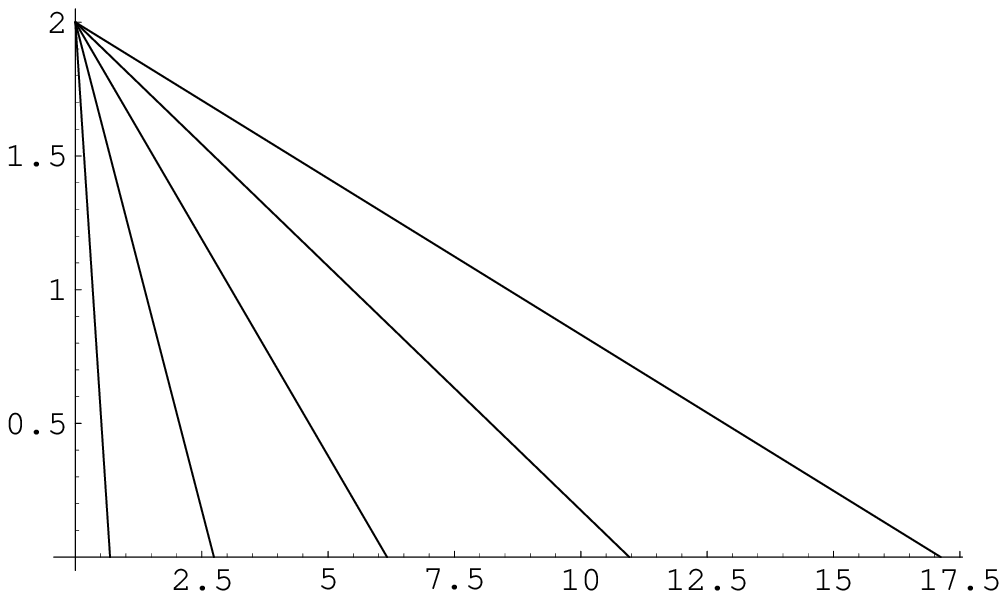}\\
\hspace{3.0cm}$T_A \rightarrow$ keV
 \caption{The function $f_{coh}(T_A)$, obtained  from $F_{coh}(T_A)$ after setting the form factor
 and the quenching factor equal to unity,
 for A=131 on the left and A=32 on the right as a function of the recoil energy ( for  neutrino energies 
3, 6, 9, 12 and 15 keV increasing to the right). 
 According to Eq.  (\ref{Eq:ymax}) the maximum recoil energy is increasing as the neutrino energy increases.} 
 \label{fig:fcoh131}
 \end{center}
   \end{figure}

  By setting the form factor
equal to unity in $F_{coh}$ we obtain $f_{coh}$. The latter
is shown in Fig. \ref{fig:fcoh131}. Note that the maximum recoil energy for boron
neutrinos cannot exceed the 4 keV in the case of Xe and 17 keV in the case of S. The differential
cross section is shown in Figs \ref{fig:dsigma131} and \ref{fig:dsigma32}. Note the rapid fall of the cross section with recoil energy. The cross section has been almost completely depleted beyond energies 1 and 5 keV for a 
intermediate ($^{131}$Xe ) and light ($^{32}$S) targets respectively.
\section{The event rate}
\begin{figure}[!ht]
 \begin{center}
 \rotatebox{90}{\hspace{1.0cm} {$f_{\nu}(E_{\nu})\rightarrow $}}
\includegraphics[scale=0.8]{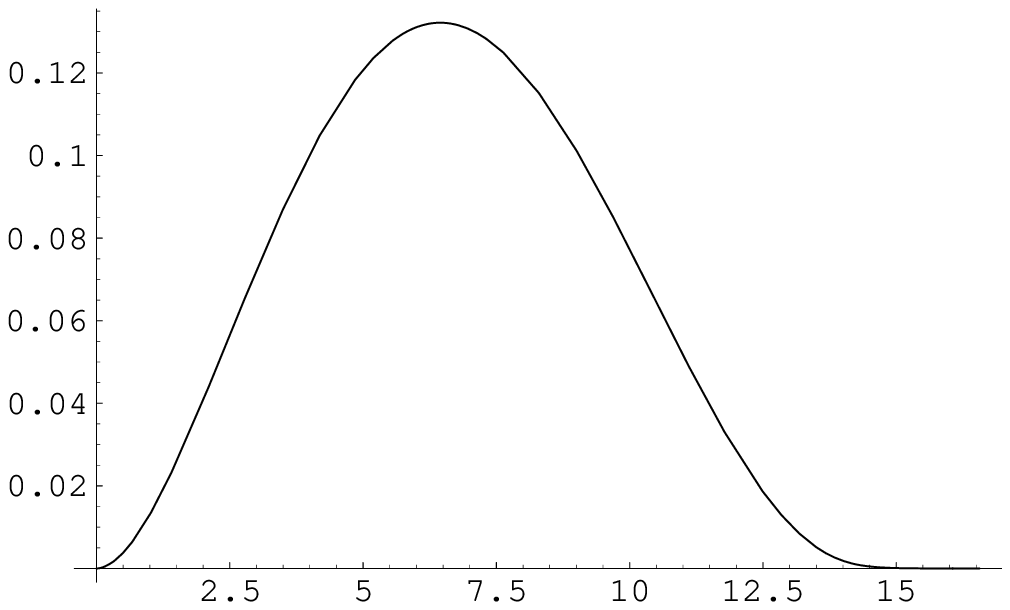}
\hspace{8.0cm}$E_{\nu} \rightarrow$ MeV
 \caption{The boron solar neutrino spectrum.} 
 \label{fig:nuspec}
 \end{center}
  \end{figure}

\begin{figure}[!ht]
 \begin{center}
 \rotatebox{90}{\hspace{1.0cm} {$\Phi_{\nu}(E_{\nu})\rightarrow cm^{-2}s^{-1}$}}
\includegraphics[scale=0.8]{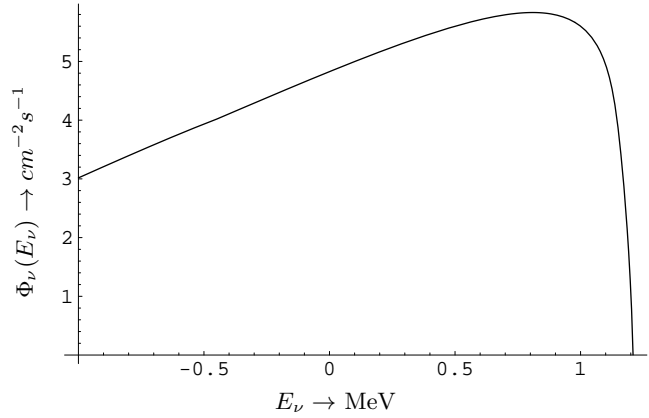}
\hspace{8.0cm}$E_{\nu} \rightarrow$ MeV
 \caption{The boron solar neutrino flux in units of $cm^{-2}s^{-1}$ in a Log-Log plot.} 
 \label{fig:nuflux}
 \end{center}
  \end{figure}

\begin{figure}[!ht]
 \begin{center}
 \rotatebox{90}{\hspace{1.0cm} {$\frac{d\sigma}{dT_A} \rightarrow $ cm$^2$ keV$^{-1}$ }}
\includegraphics[scale=1.1]{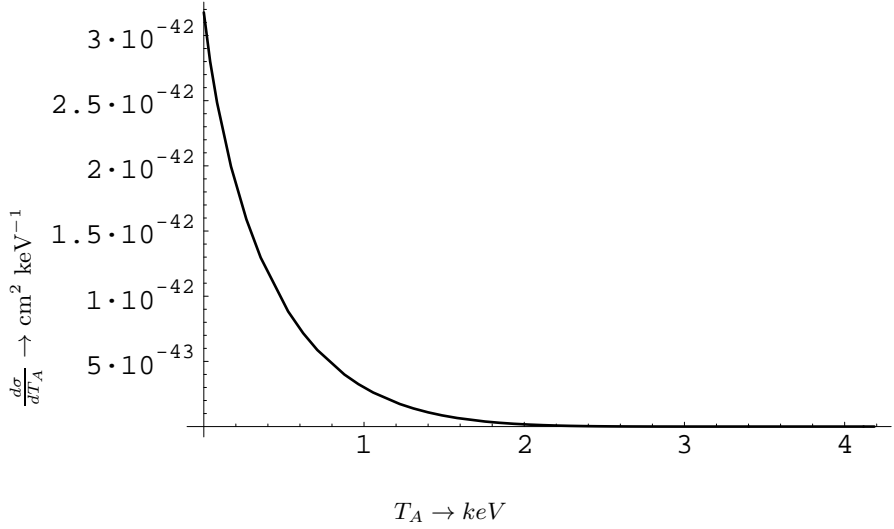}
\hspace{8.0cm}$T_A \rightarrow keV$ 
 \caption{The neutrino induced differential cross section in units cm$^2$ keV$^{-1}$  
as a function of the recoil energy in keV
in the case of the target $^{131}$Xe. Note that essentially all the contribution comes from the
 recoil energy region $0\leq T_A\leq 1$ keV.
The effect of the form factor is invisible in the figure.} 
 \label{fig:dsigma131}
 \end{center}
  \end{figure}

\begin{figure}[!ht]
 \begin{center}
 \rotatebox{90}{\hspace{1.0cm} {$\frac{d\sigma}{dT_A} \rightarrow $ cm$^2$ keV$^{-1}$ }}
\includegraphics[scale=1.1]{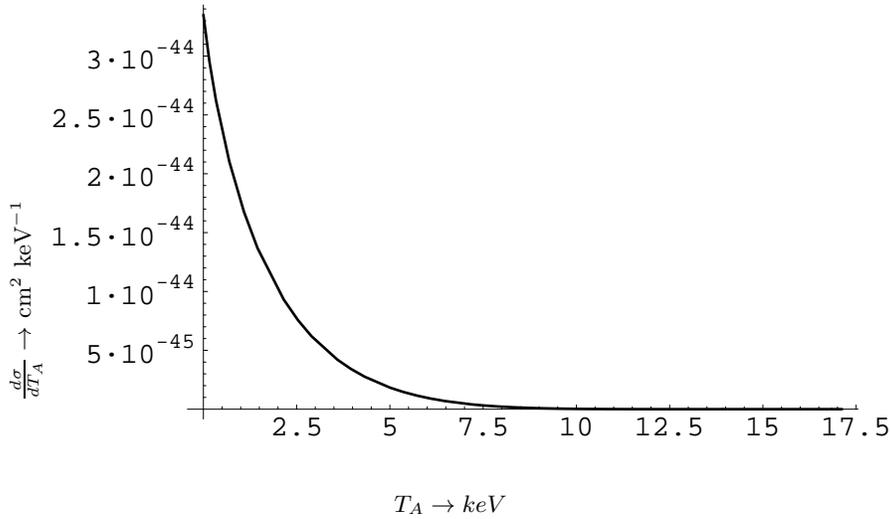}
\hspace{8.0cm}$T_A \rightarrow keV$ 
 \caption{The same as in Fig.  \ref{fig:dsigma131} for the target $^{32}$S. In this case
the recoil energy region is  wider $0\leq T_A\leq 5$ keV, but still small.}
 \label{fig:dsigma32}
 \end{center}
  \end{figure}

  To proceed further we must convolute the cross section with the neutrino spectrum. From the  neutrinos emitted by the sun only the boron neutrinos have high enough flux with sufficiently high energy to lead to nuclear recoils, which could become relevant in dark matter searches. The normalized boron neutrino spectrum is shown in
  Fig. \ref{fig:nuspec}. The corresponding flux is shown in Fig. \ref{fig:nuflux}.
The obtained differential cross section is shown in Figs  \ref{fig:dsigma131} and  \ref{fig:dsigma32}.

  Integrating the differential cross section down to zero threshold  we find the  event rates given in table
\ref{table.nurates}. The event rates are almost two orders of magnitude smaller than the rates 
for WIMP detection obtained with a nucleon cross section of $10^{-9}$pb. Thus such neutrinos cannot be a serious background for WIMP searches in the region $10^{-9}-10^{-10}$pb. In any event, as we will see below, the neutrino induced recoils are less of a background problem 
in the realistic case of non zero energy threshold.\\
Sometimes for experimental purposes one may have to focus on 
 a resticted region  of the recoill energy spectrum.
To be specific let us consider a typical low recoil energy region, e.g.   2 keV$\leq T_A\leq $4 keV.
 Clearly from Fig. \ref{fig:dsigma131} one can see that the neutrino
background is very small in this energy region. We find that the  WIMP event rate in this restricted energy region is substantially reduced, 
but it is not suppressed as much as the neutrino rate (see table \ref{table.nurates}).

\begin{table}[t]
\caption{ Comparison of the event rates for boron solar  neutrino detection with those of WIMP 
detection rates. In evaluating the latter we assumed a nucleon cross section independent of the mass. The
kinematics were obtained assuming two WIMP masses, namely 100 and 300 GeV. NoFF means that the nuclear form factor 
was neglected.
}
\label{table.nurates}
\begin{center}
\begin{tabular}{|c|c|c|c|c|c|}
\hline
 &   &  &  & &\\
&target &$R_{\chi}$(kg-y)$\times \frac{\sigma_N}{10^{-9}pb}$&$R_{\chi}$(kg-y)$\times  \frac{\sigma_N}{10^{-9}pb}$&$R_{\nu}$(kg-y)&$R_{\nu}$(kg-y);NoFF\\
\hline
&&$m_{\chi}=100$GeV&$m_{\chi}=300$GeV& \\
\hline
\rotatebox{90}{\hspace{0.0cm} {range}}&$^{131}$Xe&0.167&0.060&0.934$\times 10^{-3}$&0.952$\times 10^{-3}$\\
\rotatebox{90}{\hspace{0.0cm} { full}}&$^{32}$S&0.033&0.014&0.167$\times 10^{-3}$&0.168$\times 10^{-3}$\\
\hline
\rotatebox{90}{\hspace{0.0cm} {$\leq$ 4keV}}&$^{131}$Xe&0.018&0.010&0.308$\times 10^{-5}$&0.310$\times 10^{-5}$\\
\rotatebox{90}{\hspace{0.0cm} {2keV$\leq T_A$}}&$^{32}$S&0.0012&0.0006&0.367$\times 10^{-4}$&0.368$\times 10^{-4}$\\
\hline
\end{tabular}
\end{center}
\end{table}

We should mention that the obtained rates are independent of the neutrino oscillation parameters, since the neutral
current events, which we considered in the present calculation, are not affected by such oscillations.
\section{Quenching factors and Energy thresholds }.
The above results refer to an ideal detector operating down to zero energy threshold. For a real detector, however,
as we have already mentioned, the nuclear recoil events are quenched, especially at low energies. 
The quenching
factor for a given detector  is the ratio of the signal height for a recoil track divided by that
 of an electron signal height with the same energy. We should not forget that the signal heights depend on the
velocity and how the signals are extracted experimentally. The actual quenching
  factors must be determined experimentally for each target. In the case of NaI the quenching
factor is 0.05, while for Ge and Si it is 0.2-0.3. For our purposes it is adequate, to multiply
the energy scale by an recoil energy dependent quenching factor, $ Q_{fac}(T_A)$ 
  adequately described by the Lidhard theory \cite{LIDHART}.  More specifically in our estimate of $Qu(T_A)$ we assumed a quenching factor of the following empirical form \cite{SIMON03}-\cite{LIDHART}:
\beq
Q_{fac}(T_A)=r_1\left[ \frac{T_A}{1keV}\right]^{r_2},~~r_1\simeq 0.256~~,~~r_2\simeq 0.153 
\label{quench1}
\eeq 
\begin{figure}[!ht]
 \begin{center}
 \rotatebox{90}{\hspace{1.0cm} {$Q_{fac}(T_A)\rightarrow $}}
\includegraphics[scale=0.8]{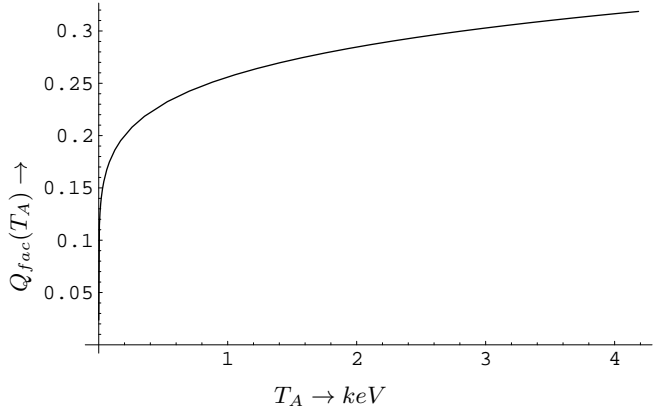}
\hspace{8.0cm}${T_A} \rightarrow keV$ 
 \caption{The quenching factor in the case of A=131 as a function of the recoil energy.} 
 \label{fig:quench131}
 \end{center}
  \end{figure}
  \begin{figure}[!ht]
 \begin{center}
 \rotatebox{90}{\hspace{1.0cm} {$Q_{fac}(T_A)\rightarrow $}}
\includegraphics[scale=0.8]{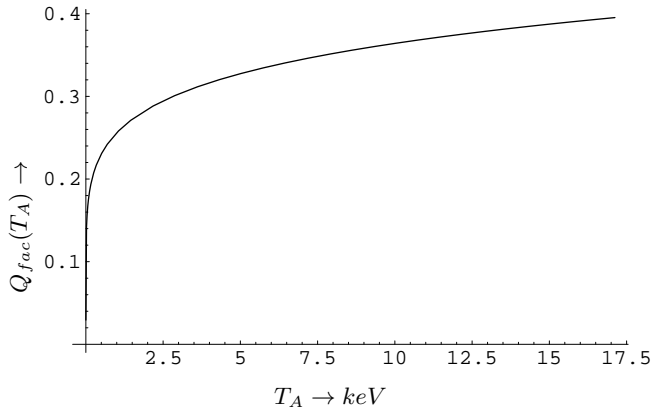}
\hspace{8.0cm}${T_A} \rightarrow keV$ 
 \caption{The quenching factor in the case of A=32 as a function of the recoil energy. It is similar to Fig. \ref{fig:quench131} except that the allowed recoil energy is different.} 
 \label{fig:quench32}
 \end{center}
  \end{figure}
 
The quenching factors 
very much depend on the detector type. 
The quenching factor,exhibited in Figs \ref{fig:quench131} and \ref{fig:quench32} for recoil energies of $^{131}$Xe
and $^{32}$S respectively, were obtained assuming the same quenching of the form of Eq. (\ref{quench1}).
In the presence of the quenching factor as given by Eq.( \ref{quench1})
the measured recoil energy is typically reduced by factors of about 3, when compared with the electron energy. In other words 
a threshold energy of electrons of 1 keV becomes 3 keV for nuclear recoils. Accordingly, the event rates
for neutrino recoils are reduced much, as seen from Figs \ref{fig:thr131} and \ref{fig:thr32} below. On the other hand
the WIMP recoil events are not reduced much, since the recoil energy is well above threshold.
%
   \\The above rates were obtained in the case of zero threshold. Due to the relatively low recoil energies, however, the 
effect of threshold is crucial (see Figs  \ref{fig:thr131} and  \ref{fig:thr32}). One clearly sees that
 the observed events are an order of magnitude down  if the energy threshold is 1 keV (2 keV) for Xe(S) 
respectively.  Thus with quenching most
  signals get below the threshold energy of $\approx$ 1 keV.
On the other hand the WIMP 
event rates are almost unaffected, unless the threshold energy becomes larger than 5 keV.
           \begin{figure}[!ht]
 \begin{center}
 \rotatebox{90}{\hspace{1.0cm} {$\frac{\sigma(E_{thr})}{\sigma(E_{thr}=0)} \rightarrow $}}
\includegraphics[scale=1.0]{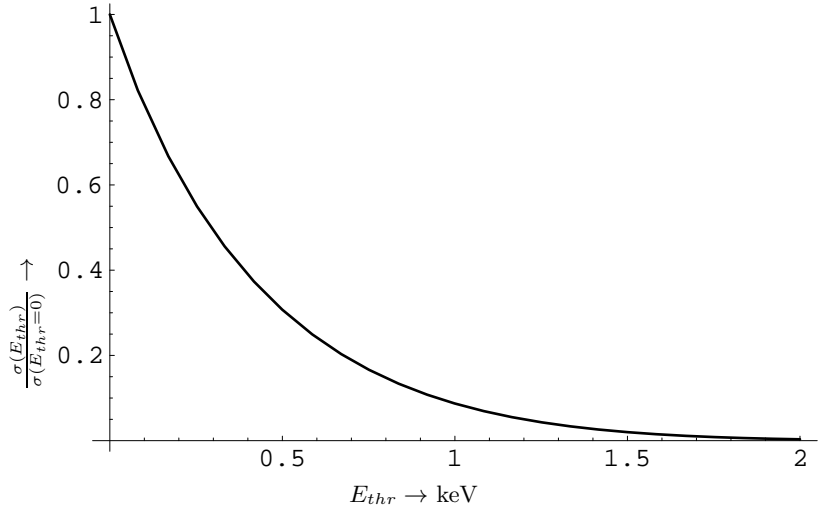}
\hspace{4.0cm}$ E_{thr}\rightarrow $ keV 
 \caption{The ratio of the total cross section with threshold divided by that with zero threshold   for A=131 as a function of the threshold energy. Otherwise the notation is similar to that of Fig. \ref{fig:dsigma131}.
Note  that
 the observed events are an order of magnitude down,  if the energy threshold is 1 keV. }
 \label{fig:thr131}
 \end{center}
  \end{figure}
          \begin{figure}[!ht]
 \begin{center}
 \rotatebox{90}{\hspace{1.0cm} {$\frac{\sigma(E_{thr})}{\sigma(E_{thr}=0)} \rightarrow $}}
\includegraphics[scale=1.0]{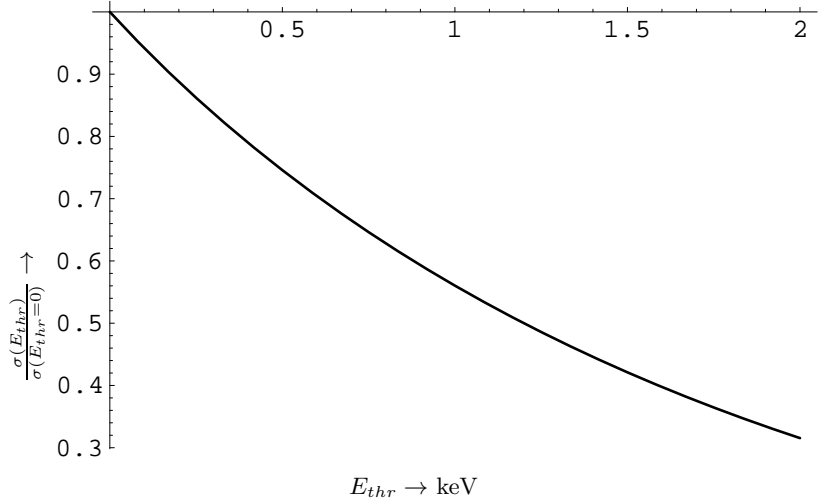}
\hspace{4.0cm}$  E_{thr}\rightarrow$ keV 
 \caption{The ratio of the total cross section with threshold divided by that with zero threshold   for A=32 as a function of the threshold energy. Otherwise the notation is similar to that of Fig. \ref{fig:thr131}. 
Note  that
 the observed events are an order of magnitude down,  if the energy threshold is above 2 keV. }
 \label{fig:thr32}
 \end{center}
  \end{figure}
  \section{concluding remarks}
  In the present study we considered the elastic scattering of WIMP-nucleus interaction and the 
corresponding elastic scattering of boron solar neutrinos. The former are favored by an A$^2$ enhancement
due to coherence of all nucleons, while the latter by N$^2$ due to the neutron coherence resulting from
the neutral current interaction. The latter may become a source of background, if the WIMP nucleon interaction
turns out to be very small. Our results can be summarized as follows:
\begin{enumerate}
\item The differential cross section for solar neutrinos decreases sharply as the nuclear recoil energy increases. It almost vanishes beyond 1 keV (5 keV) for intermediate (light target), like $^{131}Xe$ ($^{32}S$). On the other
the corresponding event rates for WIMPs of mass $\approx$ 100 GeV extend further than 30 (150) keV $^{131}Xe$ ($^{32}S$) respectively. 
\item The event rates for boron solar neutrinos at zero threshold energy and no quenching are 2-3 
orders of magnitude smaller than those for WIMPs with a nucleon cross section $10^{-9}$pb. Thus
solar neutrinos are not a serious background down to $10^{-10}$pb, but they may have to be considered
at the level of $10^{-11}$pb.
\item Since the nuclear recoil energy arising from solar neutrino scattering is smaller than that 
associated with heavy WIMPs, one  can further
substantially decrease its contribution by restricting the observation of  the recoil energy spectra
 above a few keV without seriously affecting the corresponding WIMP rates. 
Thus neutrinos do not appear to be a serious background even at the level of $10^{-11}$pb.
\item By exploiting the quenching factors one may reduce this background still further.
\item It should be noted that the solar neutrinos do not affect the DAMA result, {\it Bernabei et al} 
 \cite{CDMALL}, in both the energy and the cross section. DAMA uses the NaI target with a large quenching factor. 
 Since events from NaI are mostly due to $^{127}$I, the event rate and the quenching factor are nearly the same as those for $^{131}$Xe discussed in the text. 
Thus the solar neutrino events  are  below 1 keV, i.e. they are below the DAMA energy bin 2-4 keV.
 In other words the solar neutrinos may be dangerous for WIMP detection  for WIMP-nucleon cross section  less than $10^{-10}$ pb, which is far below the DAMA  region of $10^{-5} - 10^{-6}$ pb.
\item The observation of the annual modulation of the signal (see e.g. \cite{JDV03} and references there in)  or even better by performing
 directional experiments \cite{DRIFT},\cite{VF07},
 i.e experiments  in which the direction of recoil is also measured, one will be able to select WIMP signals
and discriminate against neutrino scattering.  
\end{enumerate}
 In the above discussion we focused on the coherent WIMP-nucleus scattering. We should not, of course, forget the
 spin contribution due to the axial current. In this case one has to deal with the proton and neutron
 spin nuclear matrix elements and the relevant elementary proton and neutron spin amplitudes. So the obtained results
 will depend on the specific target. It is clear, however, that the spin matrix elements do not exhibit
coherence, i.e.  do not scale with  A$^2$. Thus the event rates will be suppressed. In other words
the boron neutrinos maybe be
 a serious background for nucleon spin cross section at the level of 10$^{-8}$pb. This, of course, will be
 worrisome, but then the corresponding coherent cross section must be  less than 10$^{-11}$pb, 
since it is only then that the two modes can compete.
 \section{Acknowledgments}
 One of us (JDV) is indebted to professor Pantelis Kelires and the Cyprus Technical University for their
 hospitality and support. Partial financial support by MRTN-CT-2004-503369 and MRTN-CT-2006-035863 is also acknowledged. The other author (H.E.) thanks the NIRS directors and the NIRS colleagues for their support 
 and hospitality at NIRS.    
    
\end{document}